\documentclass[prd,aps,showpacs,nofootinbib,preprint,eqsecnum]{revtex4-1}
\usepackage{graphicx,color,amsmath,amsxtra}
\usepackage{epsf}
\usepackage{amssymb}
\usepackage{enumerate}
\usepackage{hhline}
\usepackage{array}
\usepackage{tabularx}

\begin{document}

\title{Observational Constraints on Exponential Gravity}

\author{Louis Yang}
\email{louis.lineage@msa.hinet.net}

\author{Chung-Chi Lee}
\email{g9522545@oz.nthu.edu.tw}

\author{Ling-Wei Luo}
\email{d9622508@oz.nthu.edu.tw}

\author{Chao-Qiang Geng}
\email{geng@phys.nthu.edu.tw}

\affiliation{Department of Physics, National Tsing Hua University, Hsinchu 300,
Taiwan}
\date{\today}

\begin{abstract}
We study the observational constraints on the exponential gravity
model of
$f(R)=-\beta R_{s}(1-e^{-R/R_{s}})$.
We use the latest observational data including Supernova Cosmology
Project (SCP) Union2 compilation, Two-Degree Field Galaxy Redshift
Survey (2dFGRS),  Sloan Digital Sky Survey Data Release 7 (SDSS
DR7) and Seven-Year Wilkinson Microwave Anisotropy Probe (WMAP7) in
our analysis. From these observations, we obtain a lower bound on the
model parameter $\beta$ at 1.27 (95\% CL) but no appreciable upper
bound. The constraint on the present matter density parameter is $0.245<\Omega_{m}^{0}<0.311$
(95\% CL). We also find out the best-fit value of model parameters
on several cases.
\end{abstract}


\maketitle

\section{Introduction}

Cosmic observations from type Ia supernovae (SNe Ia) \citep{Riess1998a,Perlmutter1999a},
large scale structure (LSS) \citep{Tegmark2004a,Seljak2005a}, baryon acoustic
oscillations (BAO) \citep{Eisenstein2005} and cosmic microwave background (CMB)
\citep{Spergel2003,Spergel2007} indicate that our universe is undergoing
an accelerating expansion. The reason for this acceleration, the so-called
dark energy problem, remains a fascinating question today. The simplest
model to explain this problem is the $\Lambda$CDM model, in which  a time
independent energy density is added to the universe. However, the
$\Lambda$CDM
model suffers from both fine-tuning and coincidence problems \citep{Weinberg1989,Sahni2000,Carroll2001,Peebles2003,Padmanabhan2003b,Copeland2006}.
In general, the ways to understand the cosmic acceleration can be separated
into two branches. One is to modify the matter by introducing some
kind of \textquotedblleft{}dark energy\textquotedblright{}. The other one
is to modify Einstein\textquoteright{}s general relativity \textendash{}
the modification of gravity.

In modified gravity, one of the popular approaches is to promote the Ricci scalar
$R$ in the Einstein-Hibert action to a function, $f(R)$.
Although there are several viable
$f(R)$ models, many of them are restricted to the regimes to be
 effectively identical to the $\Lambda$CDM by the observational
constraints.
Recently, Linder \citep{Linder2009}
has explored
an $f(R)$ theory named \textquotedblleft{}exponential gravity\textquotedblright{},
which has also been discussed in Refs. \citep{Zhang2006,Zhang2007,Cognola2008}.
The exponential gravity has the feature that it allows the relaxation
of fine-tuning and it has only one more parameter than the $\Lambda$CDM
model. In addition, the exponential gravity satisfies all  conditions for
the viability \citep{Bamba2010} such as the local gravity constraint,
stability of the late-time de Sitter point, constraints from the violation
of the equivalence principle, stability of cosmological perturbations,
positivity of the effective gravitational coupling, and asymptotic
behavior to the $\Lambda$CDM model in the high curvature regime.
In this paper, we will study the constraints given by latest observational
data, reexamine the alleviation of the fine-tuning problem, and find the
possibility of the derivation from $\Lambda$CDM.
We use units of $k_\mathrm{B} = c = \hbar = 1$ and
the gravitational constant is given by $G = M_{\mathrm{Pl}}^{-2}$
with the Planck mass of
$M_{\mathrm{Pl}} = 1.2 \times 10^{19}$\,\,GeV.

The paper is organized as follows.
In Sec. \ref{sec:Exponential-Gravity},
we review  equations of motion and the
asymptotic behavior at the high redshift regime in
the exponential gravity model. In Sec. \ref{sec:Observational-Constraints},
we discuss the observations and methods.
We show our results in Sec. \ref{sec:Results}.
Finally, conclusions are given in Sec. \ref{sec:Conclusion}.

\section{Exponential Gravity\label{sec:Exponential-Gravity}}

The action of $f(R)$ gravity with matter is given by
\begin{equation}
S=\frac{1}{2\kappa^{2}}\int d^{4}x\sqrt{-g}\left[R+f(R)\right]+S_{m},
\label{action}
\end{equation}
 where $\kappa^{2}\equiv8\pi G$ and $f(R)$ is a function of the
Ricci scalar curvature $R$. In this paper, we focus on the exponential
gravity model \citep{Linder2009}, given by
\begin{equation}
f(R)=-\beta R_{s}(1-e^{-R/R_{s}}),
\label{fR}
\end{equation}
where
$R_{s}$ is related to the characteristic curvature modification scale.
Since the product of
$\beta$ and $R_{s}$ can be determined by
the present matter density  $\Omega_{m}^{0}$~\citep{Linder2009},
we can choose $\beta$ and $\Omega_{m}^{0}$ as the free parameters in the model.

We use the standard metric formalism. From the action \eqref{action},
the modified Friedmann equation of motion becomes \citep{Song2007}
\begin{equation}
H^{2}=\frac{\kappa^{2}\rho_{M}}{3}+\frac{1}{6}(f_{R}R-f)-H^{2}\left(f_{R}+f_{RR}R'\right),\label{modified Friedmann equation}\end{equation}
 where $H\equiv\dot{a}/a$ is the Hubble parameter, a subscript R
denotes the derivative with respect to R, a prime represents $d/d\ln a$,
and $\rho_{M}=\rho_{m}+\rho_{r}$ is the energy density of all perfect
fluids of generic matter including (non-relativistic) matter, denoted
by $m$, and relativistic particles, denoted by $r$. Here, we only
consider the matter density. Since the modification by the exponential
gravity only happens at the low redshift, the contributions from relativistic
particles are negligible. In a flat spacetime, the Ricci scalar is
given by
\begin{equation}
\nonumber
R=12H^{2}+6HH'.\label{Ricci scalar curvature}
\end{equation}

Following Hu and Sawicki's parameterization \citep{Hu2007}, we define

\begin{equation}
y_{H}\equiv\frac{\rho_{DE}}{\rho_{m}^{0}}=\frac{H^{2}}{m^{2}}-a^{-3},\quad y_{R}\equiv\frac{R}{m^{2}}-3a^{-3},
\label{parameters}
\end{equation}
 where $m^{2}\equiv\kappa^{2}\rho_{m}^{0}/3$, $\rho_{DE}$ is the effective
dark energy density, and $\rho_{m}^{0}$ is the present matter density.
Then,  Eqs.~\eqref{modified Friedmann equation} and \eqref{Ricci scalar curvature}
can be rewritten as two coupled differential equations,
\begin{equation}
y_{H}'=\frac{y_{R}}{3}-4y_{H}\label{yH'}
\end{equation}
 and

\begin{equation}
y_{R}'=9a^{-3}-\frac{1}{H^{2}f_{RR}}\left[y_{H}+f_{R}\left(\frac{H^{2}}{m^{2}}-\frac{R}{6m^{2}}\right)+\frac{f}{6m^{2}}\right],\label{yR'}
\end{equation}
 where $R$ and $H^{2}$ can be further replaced by $y_{R}$ and $y_{H}$
from equations in \eqref{parameters}. Combining Eqs.~\eqref{yH'}
and \eqref{yR'}, we obtain a second order differential equation of
$y_{H}$,
\begin{equation}
y_{H}''+J_{1}y_{H}'+J_{2}y_{H}+J_{3}=0,\label{2nd diff eq}
\end{equation}
 where
 \begin{eqnarray}
J_{1}&=&4-\frac{1}{y_{H}+a^{-3}}\frac{f_{R}}{6m^{2}f_{RR}},\nonumber\\
J_{2}&=&-\frac{1}{y_{H}+a^{-3}}\frac{f_{R}-1}{3m^{2}f_{RR}},\nonumber\\
J_{3}&=&-3a^{-3}+\frac{f_{R}a^{-3}+f/3m^{2}}{y_{H}+a^{-3}}\frac{1}{6m^{2}f_{RR}},
\end{eqnarray}
 with
 \begin{eqnarray}
R=m^{2}\left[3\left(y_{H}'+4y_{H}\right)+3a^{-3}\right].
\end{eqnarray}
Solving Eq.~\eqref{2nd diff eq}  numerically,  we
can get the evolution of the Hubble parameter in the low redshift regime
($z=0\sim4$). The effective dark energy equation of state $w_{DE}$
is given by
\begin{eqnarray}
w_{DE}=-1-\frac{y_{H}'}{3y_{H}}.
\end{eqnarray}

In the high redshift regime ($z\gtrsim4$), the exponential factor $e^{-R/R_{S}}$ of $f(R)$ in Eq.~(\ref{fR}) becomes negligible ($e^{-R/R_{S}}<10^{-5}$).
The exponential gravity model behaves essentially like a cosmological
constant model with the dark energy density parameter
$\Omega_{\Lambda}=\beta R_{S}/6H_{0}^{2}\cong\Omega_{m}^{0}y_{H}(z_{high})$. 
Thus, the Hubble parameter as a function of $z$ in this regime can be expressed as
\begin{eqnarray}
H(z)&=&H_{0}\sqrt{\Omega_{m}^{0}\left(1+z\right)^{3}+\Omega_{r}^{0}\left(1+z\right)^{4}+\frac{\beta R_{S}}{6H_{0}^{2}}},
\label{eqn2-11}
\end{eqnarray}
 where $\Omega_{r}^{0}$ is the density parameter of relativistic particles
 including photons and neutrinos\footnote{$\Omega_{r}^{0}=\Omega_{\gamma}^{0}\left(1+0.2271N_{eff}\right)$,
where $\Omega_{\gamma}^{0}$ is the present fractional photon energy density
and $N_{eff}=3.04$ is the effective number of neutrino species \citep{Komatsu2010}.}.
The equation (\ref{eqn2-11}) will be used in the data fitting of CMB and the high redshift part of BAO in section \ref{sec:Observational-Constraints}.

\section{Observational Constraints\label{sec:Observational-Constraints}}

To constrain the free parameters of $\beta$ and $\Omega_{m}^{0}$ in the exponential gravity model,
we use three kinds of the observational data including
SNe Ia, BAO and CMB. The SNe Ia and CMB data lead to constraints at the low and high redshift regimes, respectively, while the BAO data provide constraints at the both regimes.

\subsection{Type Ia Supernovae (SNe Ia)}

The observations of SNe Ia, known as {}``standard candles'',
give us the information about the luminosity distance $D_{L}$ as a function
of the redshift $z$. The distance modulus $\mu$ is defined as
\begin{eqnarray}
\mu_{th}(z_{i})\equiv5\log_{10}D_{L}(z_{i})+\mu_{0},
\end{eqnarray}
 where $\mu_{0}\equiv42.38-5\log_{10}h$ with $H_{0}=h\cdot100km/s/Mpc$
is the present value of the Hubble parameter.
The Hubble-free luminosity
distance for the flat universe is
\begin{eqnarray}
D_{L}(z)=(1+z)\int_{0}^{z}\frac{dz'}{E(z')},
\end{eqnarray}
 where $E(z)=H(z)/H_{0}$.
 The $\chi^{2}$ of the SNe Ia data is
\begin{eqnarray}
\chi_{SN}^{2}=\sum_{i}\frac{\left[\mu_{obs}(z_{i})-\mu_{th}(z_{i})\right]^{2}}{\sigma_{i}^{2}},
\end{eqnarray}
 where $\mu_{obs}$ is the observed value of the distance modulus.
Since the absolute magnitude of SNe Ia is unknown,
we should minimize  $\chi_{SN}^{2}$ with respect to $\mu_{0}$,
which relates to the absolute magnitude,
and expand it to be
\citep{Nesseris2005,Perivolaropoulos2005}
\begin{eqnarray}
\chi_{SN}^{2}=A-2\mu_{0}B+\mu_{0}^{2}C,
\end{eqnarray}
 where
 \begin{eqnarray}
A&=&\sum_{i}\frac{\left[\mu_{obs}(z_{i})-\mu_{th}(z_{i};\mu_{0}=0)\right]^{2}}{\sigma_{i}^{2}},
\nonumber\\
B&=&\sum_{i}\frac{\mu_{obs}(z_{i})-\mu_{th}(z_{i};\mu_{0}=0)}{\sigma_{i}^{2}},\quad C=\sum_{i}\frac{1}{\sigma_{i}^{2}}.
\end{eqnarray}
The minimum of $\chi_{SN}^{2}$ with respect to $\mu_{0}$ is
\begin{eqnarray}
\tilde{\chi}_{SN}^{2}=A-\frac{B^{2}}{C}.
\end{eqnarray}
We adopt this $\tilde{\chi}_{SN}^{2}$ for our later $\chi^{2}$ minimization.
We will use the data from the Supernova
Cosmology Project (SCP) Union2 compilation, which contains 557 supernovae
\citep{Amanullah2010}, ranging from $z=0.015$ to $z=1.4$.

\subsection{Baryon Acoustic Oscillations (BAO)}

The observation of BAO
measures the distance ratios of $d_{z}\equiv r_{s}(z_{d})/D_{V}(z)$,
where $D_{V}$ is the volume-averaged distance, $r_{s}$ is the comoving
sound horizon and $z_{d}$ is the redshift at the drag epoch \citep{Percival2010}.
The volume-averaged distance $D_{V}(z)$ is defined as \citep{Eisenstein2005}
\begin{eqnarray}
D_{V}(z)\equiv\left[(1+z)^{2}D_{A}^{2}(z)\frac{z}{H(z)}\right]^{1/3},
\end{eqnarray}
 where $D_{A}(z)$ is the proper angular diameter distance:
 \begin{eqnarray}
D_{A}(z)=\frac{1}{1+z}\int_{0}^{z}\frac{dz'}{H(z')},\quad\textrm{(for flat universe)}.
\end{eqnarray}
The comoving sound horizon $r_{s}(z)$ is given by
\begin{eqnarray}
r_{s}(z)=\frac{1}{\sqrt{3}}\int_{0}^{1/(1+z)}\frac{da}{a^{2}H({\scriptstyle z'=\frac{1}{a}-1})\sqrt{1+(3\Omega_{b}^{0}/4\Omega_{\gamma}^{0})a}},
\end{eqnarray}
 where $\Omega_{b}^{0}$ and $\Omega_{\gamma}^{0}$ are the present
values of baryon and photon density parameters, respectively. We use
$\Omega_{b}^{0}=0.022765h^{-2}$ and $\Omega_{\gamma}^{0}=2.469\times10^{-5}h^{-2}$
\citep{Komatsu2010}. The fitting formula for $z_{d}$ is given by
\citep{Eisenstein1998}
\begin{eqnarray}
z_{d}=\frac{1291(\Omega_{m}^{0}h^{2})^{0.251}}{1+0.659(\Omega_{m}^{0}h^{2})^{0.828}}\left[1+b_{1}(\Omega_{b}^{0}h^{2})^{b2}\right],
\end{eqnarray}
 where
\begin{eqnarray}
b_{1}&=&0.313(\Omega_{m}^{0}h^{2})^{-0.419}\left[1+0.607(\Omega_{m}^{0}h^{2})^{0.674}\right],
\nonumber\\
b_{2}&=&0.238(\Omega_{m}^{0}h^{2})^{0.223}.
\end{eqnarray}
The typical value of $z_{d}$ is about 1021 with $\Omega_{m}^{0}=0.276$ and $h=0.705$.
Since $z_{d}$ is in the high redshift regime, we use Eq. (\ref{eqn2-11}) to calculate $r_{s}(z_{d})$.
On the other hand,
$D_{V}(z)$ is evaluated by the numerical result of Eq. (\ref{2nd diff eq}) as it is
in the low redshift regime.

The BAO data from the Two-Degree Field Galaxy Redshift Survey (2dFGRS)
and the Sloan Digital Sky Survey Data Release 7 (SDSS DR7) \citep{Percival2010}
measured the distance ratio $d_{z}$ at two redshifts $z=0.2$ and
$z=0.35$ to be $d_{z=0.2}^{obs}=0.1905\pm0.0061$ and $d_{z=0.35}^{obs}=0.1097\pm0.0036$
with the inverse covariance matrix: \begin{eqnarray}
C_{BAO}^{-1}=\left(\begin{array}{cc}
30124 & -17227\\
-17227 & 86977\end{array}\right).\end{eqnarray}
The $\chi^{2}$ for the BAO data is \begin{eqnarray}
\chi_{BAO}^{2}=(x_{i,BAO}^{th}-x_{i,BAO}^{obs})(C_{BAO}^{-1})_{ij}(x_{j,BAO}^{th}-x_{j,BAO}^{obs}),\end{eqnarray}
 where $x_{i,BAO}\equiv\left(d_{0.2},d_{0.35}\right)$.

\subsection{Cosmic Microwave Background (CMB)}

The CMB is sensitive to the distance
to the decoupling epoch $z_{*}$ \citep{Komatsu2009}. It can give
constraints on the model in the high redshift regime ($z\sim1000$).
The CMB data are taken from Wilkinson Microwave Anisotropy Probe (WMAP)
observations \citep{Komatsu2010}. To use the WMAP data, we compare
three quantities: (i) the acoustic scale $l_{A}$,
\begin{eqnarray}
l_{A}(z_{*})\equiv(1+z_{*})\frac{\pi D_{A}(z_{*})}{r_{S}(z_{*})},
\end{eqnarray}
(ii) the shift parameter $R$ \citep{Bond1997},
\begin{eqnarray}
R(z_{*})\equiv\sqrt{\Omega_{m}^{0}}H_{0}(1+z_{*})D_{A}(z_{*}),
\end{eqnarray}
and (iii) the redshift of the decoupling epoch $z_{*}$. The fitting
function of $z_{*}$ is given by \citep{Hu1996}
\begin{eqnarray}
z_{*}=1048\left[1+0.00124(\Omega_{b}^{0}h^{2})^{-0.738}\right]\left[1+g_{1}(\Omega_{m}^{0}h^{2})^{g2}\right],
\end{eqnarray}
where
\begin{eqnarray}
g_{1}=\frac{0.0783(\Omega_{b}^{0}h^{2})^{-0.238}}{1+39.5(\Omega_{b}^{0}h^{2})^{0.763}},\quad g_{2}=\frac{0.560}{1+21.1(\Omega_{b}^{0}h^{2})^{1.81}}.
\end{eqnarray}

The $\chi^{2}$ of the CMB data is
\begin{eqnarray}
\chi_{CMB}^{2}=(x_{i,CMB}^{th}-x_{i,CMB}^{obs})(C_{CMB}^{-1})_{ij}(x_{j,CMB}^{th}-x_{j,CMB}^{obs}),
\end{eqnarray}
 where $x_{i,CMB}\equiv\left(l_{A}(z_{*}),R(z_{*}),z_{*}\right)$
and $C_{CMB}^{-1}$ is the inverse covariance matrix. The
data from Seven-Year Wilkinson Microwave Anisotropy Probe (WMAP7)
observations \citep{Komatsu2010} lead to $l_{A}(z_{*})=302.09$,
$R(z_{*})=1.725$ and $z_{*}=1091.3$ with the inverse covariance
matrix:
\begin{eqnarray}
C_{CMB}^{-1}=\left(\begin{array}{ccc}
2.305 & 29.698 & -1.333\\
29.698 & 6825.27 & -113.180\\
-1.333 & -113.180 & 3.414\end{array}\right).
\end{eqnarray}

Finally, the $\chi^{2}$ of all the observational data is
\begin{eqnarray}
\chi^{2}=\tilde{\chi}_{SN}^{2}+\chi_{BAO}^{2}+\chi_{CMB}^{2}.
\end{eqnarray}

In our fitting process, we did not use the Markov chain Monte Carlo (MCMC) approach
because the numerical calculation for each solution of $f(R)$ theory is very time-consuming,
and the necessary change to the code like CosmoMC \cite{Lewis2002} is very extensive with no obvious benefit in our study of the exponential gravity.
Therefore, we take the simple $\chi^{2}$ method as our main fitting procedure. 
The $\Lambda$CDM result obtained from SNe Ia, BAO and CMB constraints with this $\chi^{2}$ method is $\Omega_{m}^{0}=0.276_{-0.013}^{+0.014}$,
while that with the MCMC method is $\Omega_{m}^{0}=0.272_{-0.011}^{+0.013}$ \cite{Gong2010}.
We note that the fitting in Ref. \cite{Gong2010} has also included the observational constraints from the radial BAO and Hubble parameter H(z).

%
\begin{figure}[h]
\centering{}
\includegraphics[width=1\columnwidth]{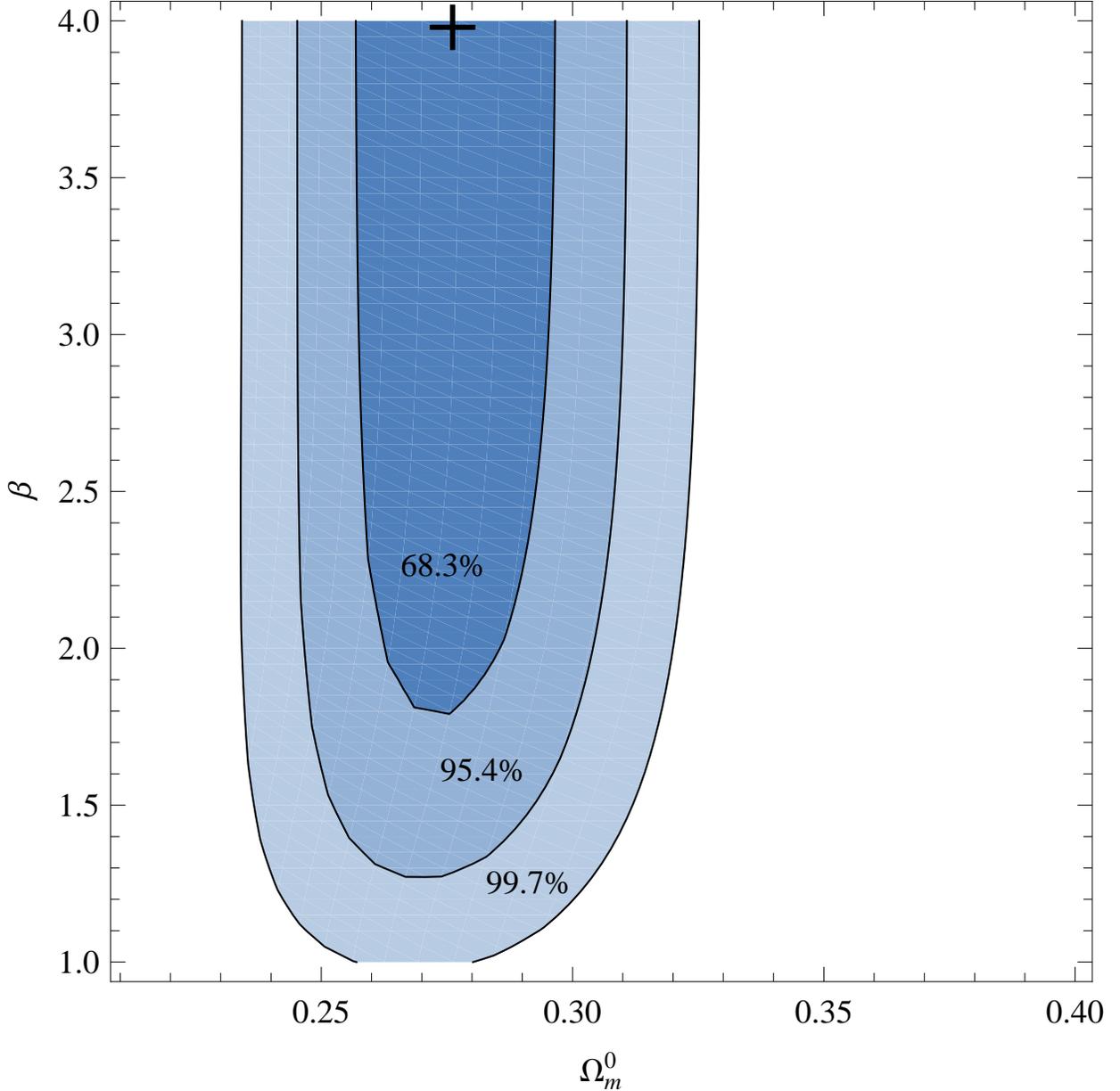}
\caption{The 68.3\%, 95.4\% and 99.7\% confidence intervals for the exponential
gravity model, constrained by the SNe Ia, BAO, and CMB data. The best-fit
point in this parameter region is marked with a plus sign.
\label{fig:contours}}

\end{figure}

\begin{table}[h]
\caption{The best-fit values of the matter density parameter $\Omega_{m}^{0}$
(68\% CL) and $\chi^{2}$ for the exponential gravity model with $\beta=2,3,4$
and the $\Lambda$CDM model. Note that the error for $\Omega_{m}^{0}$
is obtained when $\beta$ is fixed.\label{tab:comparison}}

\begin{ruledtabular}\begin{tabular}{cccc}
Model && $\Omega_{m}^{0}$ & $\chi^{2}$\tabularnewline
\hline
 & $\beta=2$ & $0.274_{-0.013}^{+0.014}$ & 546.7136\tabularnewline
Exponential Gravity & $\beta=3$ & $0.276_{-0.013}^{+0.014}$ & 545.3836\tabularnewline
 & $\beta=4$ & $0.276_{-0.013}^{+0.014}$ & 545.1721\tabularnewline
\hline
$\Lambda$CDM & & $0.276_{-0.013}^{+0.014}$ & 545.1522\tabularnewline
\end{tabular}\end{ruledtabular}
\centering{}
\end{table}

\section{Results\label{sec:Results}}

Based on the methods described in Sec. \ref{sec:Observational-Constraints},
we now examine the parameter space of the exponential gravity model. In
Fig.~\ref{fig:contours}, we present likelihood contour plots at
 68.3, 95.4 and 99.7\% confidence levels obtained from the SNe
Ia, BAO and CMB constraints. The results show that the observational
data give no upper bound on the model parameter $\beta$, making it
a free parameter. Hence, there is no fine-tuning problem. However,
a larger value of $\beta$, which is closer to the $\Lambda$CDM model,
is slightly preferred by the observational data as expected. The lower bound on
$\beta$ is $\beta>1.27$ (95\% CL). The present matter density parameter
$\Omega_{m}^{0}$ is constrained to $0.245<\Omega_{m}^{0}<0.311$
(95\% CL), which agrees with the current observations. The best-fit value
(smallest $\chi^{2}$) in the parameter space between $\beta=1$ and 4\footnote{We only concentrate on the region of  $1<\beta<4$. For $\beta>4$, it is almost the $\Lambda$CDM
model. For $\beta<1$, it is ruled out by the local gravity constraints
and the stability of the de-Sitter phase.} is $\chi^{2}=545.1721$ with $\beta=4$ and $\Omega_{m}^{0}=0.276$.
The comparison of  the best-fit $\Omega_{m}^{0}$ and $\chi^{2}$ for the model with $\beta=2,3,4$
 and $\Lambda$CDM  is shown in Table~\ref{tab:comparison}.

\begin{figure}[h]
\begin{centering}
\includegraphics[width=1\columnwidth]{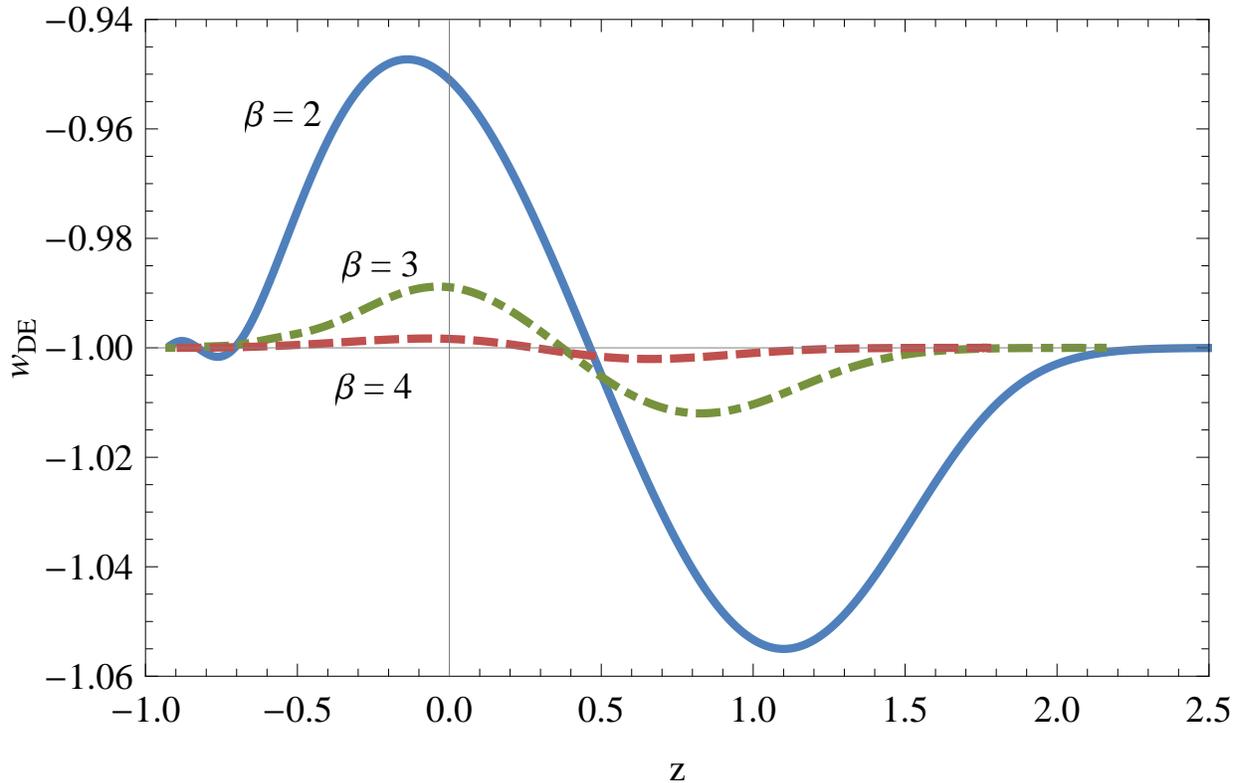}
\par\end{centering}

\caption{Evolution of the effective dark energy equation of state $w_{DE}$
corresponding to $\beta=2,3,4$ with their best-fit $\Omega_{m}^{0}$
given in Table \ref{tab:comparison}.
\label{fig:equation of state}}

\end{figure}

In Fig.~\ref{fig:equation of state}, we illustrate the evolution
of the effective dark energy equation of state $w_{DE}$ for $\beta=2,3,4$
with their best-fit $\Omega_{m}^{0}$, which is given in Table~\ref{tab:comparison}.
We can see that, for every value of $\beta$, the effective dark energy
equation of state $w_{DE}$ starts at the phase of a cosmological
constant $w_{DE}=-1$ and evolves from the phantom phase ($w_{DE}<-1$)
to the non-phantom phase ($w_{DE}>-1$). And, for larger value of
$\beta$, the deviation from cosmological constant phase ($w_{DE}=-1$)
become smaller. For $\beta=2$, there is still another small oscillation
after the main phantom phase crossing.
Negative $z$ means the future evolution.
It is clear that the exponential gravity
model has the feature of crossing the phantom phase in the past as well as the future \citep{BGL2}.

\begin{figure}[h]
\includegraphics[width=1\columnwidth]{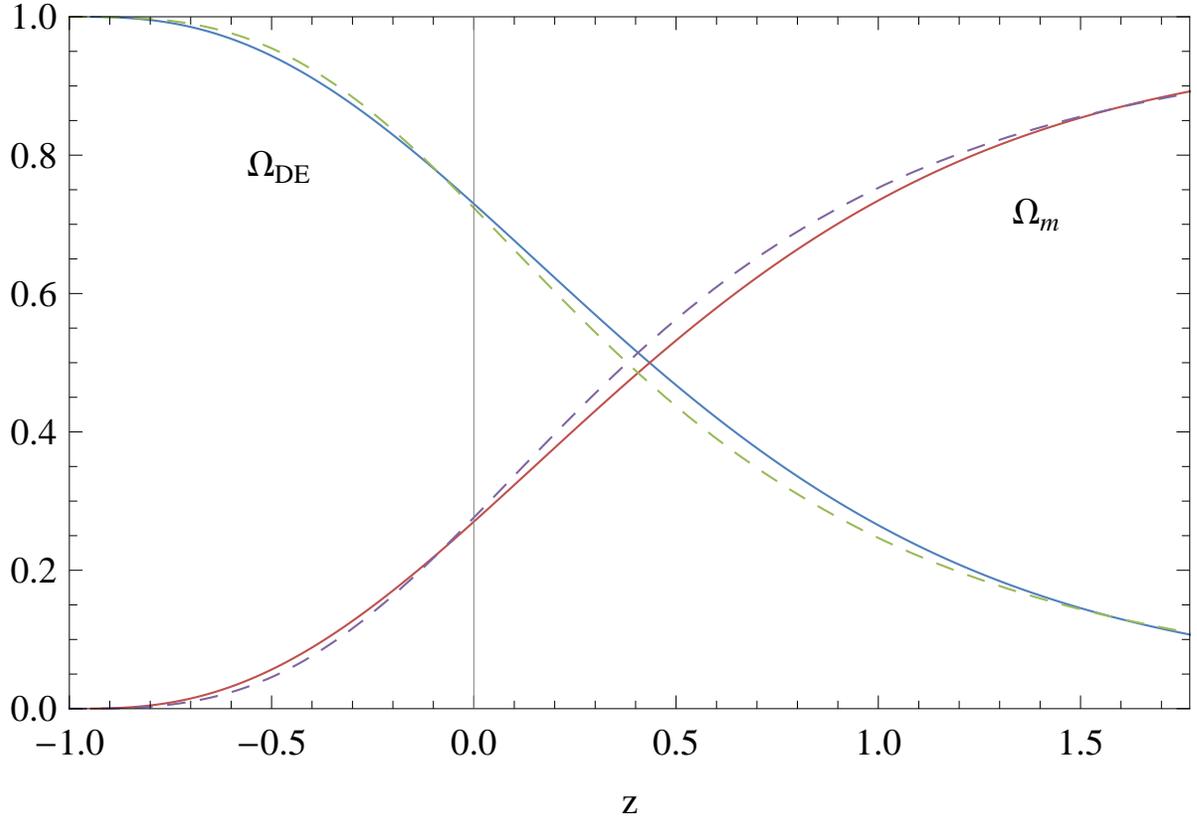}
\caption{The evolutions of the effective dark energy density parameter $\Omega_{DE}$
and non-relativistic matter density parameter $\Omega_{m}$ as functions
of $z$,
where the solid lines indicate the exponential gravity model with $\beta=1.27$ and the best-fit
 $\Omega_{m}^{0}=0.270$ and the dashed lines represent the $\Lambda$CDM model with $\Omega_{m}^{0}=0.276$.
For a higher value of $\beta$, the evolution becomes closer to that in $\Lambda$CDM.
\label{fig:density parameter}}

\end{figure}

In Fig. \ref{fig:density parameter}, we depict 
the effective dark energy density $\Omega_{DE}$
and non-relativistic matter density $\Omega_{m}$ vs. the redshift $z$.

\section{Conclusion\label{sec:Conclusion}}

We have studied the exponential gravity model. In the low redshift
regime, we follow Hu and Sawicki's parameterization to form the differential
equation for the exponential gravity and solve it numerically. In the
high redshift regime, we take advantage of the asymptotic behavior
of the exponential gravity toward an effective cosmological constant.
The analytical form of the Hubble parameter $H$ as a function of
the redshift $z$ can be expressed in the high redshift limit. We have constrained
the parameter space of the model by the SNe Ia, BAO and
CMB data. We have found that there is a lower bound on the model parameter
$\beta$ at 1.27 but no upper limit, and $\Omega_{m}^{0}$ is constrained
to the concordance value. This means that the exponential gravity model shows
no need of fine-tuning. Nevertheless, the $\Lambda$CDM model is still
included by the observational constraints since $\beta\rightarrow\infty$
corresponds to the model. Current observational data still lack the ability to
distinguish between the $\Lambda$CDM and exponential gravity models.

Finally, we remark that as seen from
Fig.~\ref{fig:density parameter},
 the noticeable difference
between the exponential gravity and $\Lambda$CDM models lies in the
regime $0.2<z<1$, and is maximized at $z=0.5$ if we compare their
expected distance modulus. An improvement on the BAO observation may
give a stronger constraint on this redshift
regime or higher.
The ongoing and future dark energy survey projects which will observe
 BAO include  WiggleZ  \cite{Glazebrook2007},  BOSS (Baryon
Oscillation Spectroscopic Survay)
\cite{BOSS}, HETDEX (Hobby-Eberly Dark Energy Experiment) \cite{HETDEX},
EUCLID \cite{Euclid},  JDEM (Joint Dark
Energy Mission)/Omega with Wide Field Infrared Survey Telescope
(WFIRST) \cite{JDEM},  BigBOSS (Big Baryon Oscillation Spec-troscopic Survay) \cite{BigBOSS},
SKA (Square Kilometer Array) \cite{SKA},
 LSST (Large Synoptic Survey Telescope) \cite{Tyson2002}
and  DES (Dark Energy Survey) \cite{DES}.
In addition, it is known that
the measurement on the growth
rate of  $f_{g}(z)=d\ln\delta_{m}/d\ln a$ has
the potential to distinguish the models with the same expansion history
but different physics. In the exponential gravity case, the growth index
is $\gamma=0.540$ for $\beta=2$. It is clear that if those surveys
 such as WiggleZ, EUCLID, BigBOSS
and JDEM/Omega can measure the growth rate  with a high accuracy, they will
be able to discriminate the exponential gravity from the $\Lambda$CDM
model.

\begin{acknowledgments}
We thank Dr. K. Bamba for
many helpful discussions and suggestions.
The work is supported in part by
the National Science Council of R.O.C. under:
Grant \#:
NSC-98-2112-M-007-008-MY3
and
National Tsing Hua University under the Boost Program \#: 97N2309F1.

\end{acknowledgments}


\end{document}